\documentclass[%
 reprint,
 amsmath,amssymb,
 aps,
]{revtex4-2}

\usepackage{graphicx}
\usepackage{dcolumn}
\usepackage{bm}
\usepackage{graphicx}
\usepackage{amsmath}
\usepackage{array}
\usepackage{multirow}

\newcommand{\etal}{\textit{et al.}}

\begin{document}

\preprint{AIP/123-QED}

\title[Advances in actinide thin films]{Advances in actinide thin films: synthesis, properties, and future directions}

\author{Kevin D. Vallejo$^{1,*}$, Firoza Kabir$^{1,2}$, Narayan Poudel$^1$, Chris A. Marianetti$^3$, David H. Hurley$^1$, Paul J. Simmonds$^{4,5}$, Cody A. Dennett$^1$, and Krzysztof Gofryk$^{1,\dag}$}
\address{$^{1}$Condensed Matter and Materials Physics, Idaho National Laboratory, Idaho Falls, ID 83415, USA}
\address{$^{2}$Glenn T. Seaborg Institute, Idaho National Laboratory, Idaho Falls, ID 83415, USA}
\address{$^{3}$Department of Applied Physics and Applied Mathematics, Columbia University, New York, NY 10027, USA}
\address{$^{4}$Department of Physics, Boise State University, Boise, ID 83725, USA}
\address{$^{5}$Micron School of Materials Science and Engineering, Boise State University, Boise, ID 83725, USA}

\date{\today}

\begin{abstract}
Actinide-based compounds exhibit unique physics due to the presence of 5\textit{f} electrons, and serve in many cases as important technological materials. Targeted thin film synthesis of actinide materials has been successful in generating high-purity specimens in which to study individual physical phenomena. These films have enabled the study of the unique electron configuration, strong mass renormalization, and nuclear decay in actinide metals and compounds. The growth of these films, as well as their thermophysical, magnetic, and topological properties, have been studied in a range of chemistries, albeit far fewer than most classes of thin film systems. This relative scarcity is the result of limited source material availability and safety constraints associated with the handling of radioactive materials. 
Here, we review recent work on the synthesis and characterization of actinide-based thin films in detail, describing both synthesis methods and modelling techniques for these materials. We review reports on pyrometallurgical, solution-based, and vapor deposition methods. We highlight the current state-of-the-art in order to construct a path forward to higher quality actinide thin films and heterostructure devices.
\end{abstract}

\maketitle

\section{Introduction}
Due to their strong electronic correlations and populations of dual-natured 5\textit{f} electrons, the actinide elements, with atomic numbers ranging from 89 to 103, represent a rather poorly understood section of the periodic table. These transition elements with partially filled 5\textit{f} electronic subshells exhibit a wide range of exotic properties. They have multiple oxidation states: in an aqueous solution, Pu can exist in four different states simultaneously, for example. Their 5\textit{f} electrons can occupy itinerant or highly localized states~\cite{morss2006chemistry}. They also exhibit divergent magnetic properties, from nearly magnetic (Np and Pu) to transition-metal-like behavior (Th, Pa, U)~\cite{mattenberger1986anisotropic}. Technologically, the most widely used actinide compound to date has been UO$_2$, in its role as the primary fuel for commercial nuclear energy production~\cite{enriquez2020structural,hurley2021thermal}.

Despite these unique physical properties, their scarcity, inherent radioactivity, and challenges concerning their synthesis mean that actinide materials are still some of the least understood elements. Transuranic actinides and their compounds have been confined to specialized facilities where materials control and accountability can be ensured. Their study and understanding are further hindered by their low natural abundance and typically short half-lives, which limit material supply~\cite{morss2006chemistry}. Developing an understanding of the fundamental properties and performance characteristics of actinide compounds is hence a challenging but highly active area of research~\cite{rickert2021hydrothermal}. Understanding the physics of these complex materials more deeply will ensure their continued technological relevance, and open new avenues for discovery in the domains of quantum science and strongly correlated systems.  

One way to enable this expansion in our understanding is to minimize the structural complexity in experimental specimens. In practice, this could include removing crystallographic defects, grain boundaries, chemical heterogeneities, and other imperfections during the synthesis process. To this end, thin film synthesis techniques have been used to produce high-quality, single crystalline specimens of various actinide compounds, including those used in nuclear fuel to study their physical properties~\cite{gouder1998thin}. Such efforts also aid comparison between experimental measurements and results from modeling and simulation tools. The geometries of these thin-film samples also permit the implementation of ``microlaboratory'' experimental designs, wherein chemical reactions may be explored in detail~\cite{Mudiyanselage_2019}. It is important to note that due to their scarcity, reactivity, radioactivity, and toxicity, high-purity actinide samples in \emph{any} geometry have only recently become available to use as targets or sources~\cite{spirlet1982preparation}. However, despite these constraints, researchers have found ways to explore a range of physical, chemical, and solution-based techniques capable of producing actinide-based films with varying degrees of success and quality. Here we catalog the successes, 
and highlight some of the unique property and performance data generated to date.

This review is organized as follows: in section \ref{sec:synthesis} we describe the synthesis methods used to create actinide-based thin films. Section \ref{sec:actinide} provides a review of actinide thin films that have been successfully synthesized, along with the techniques used for their structural, electronic, and surface characterization. Section \ref{sec:properties} summarizes the current and theorized properties of actinide thin films, and in section \ref{sec:modelling} we give an overview of the efforts and challenges related to modelling these structures. Finally, in section \ref{sec:future} we explore the rich scientific and technological promise of combining different actinide thin films into heterostructures, together with the associated challenges for this next step in the use of these important compounds.

\section{Synthesis Methods}\label{sec:synthesis}
Ever since the discovery of radiation and its properties, there has been a need to refine and purify actinide materials. During the twentieth century, the focus shifted from Ra to U in the more than 180 minerals containing these elements~\cite{edwards2000uranium}. Motivated by the applications of radioactive compounds, technology evolved and reports began to emerge of the synthesis of actinide single crystals and thin films. One of the earliest examples is from 1955, when Pu and U were dissolved in a cellulose lacquer and painted on metallic foils~\cite{glover1955method}. The intervening 70 years have seen synthesis techniques developed that incorporate the refined safety measures necessary to handle these materials.

\begin{figure*}
    \centering
    \includegraphics[width=\textwidth]{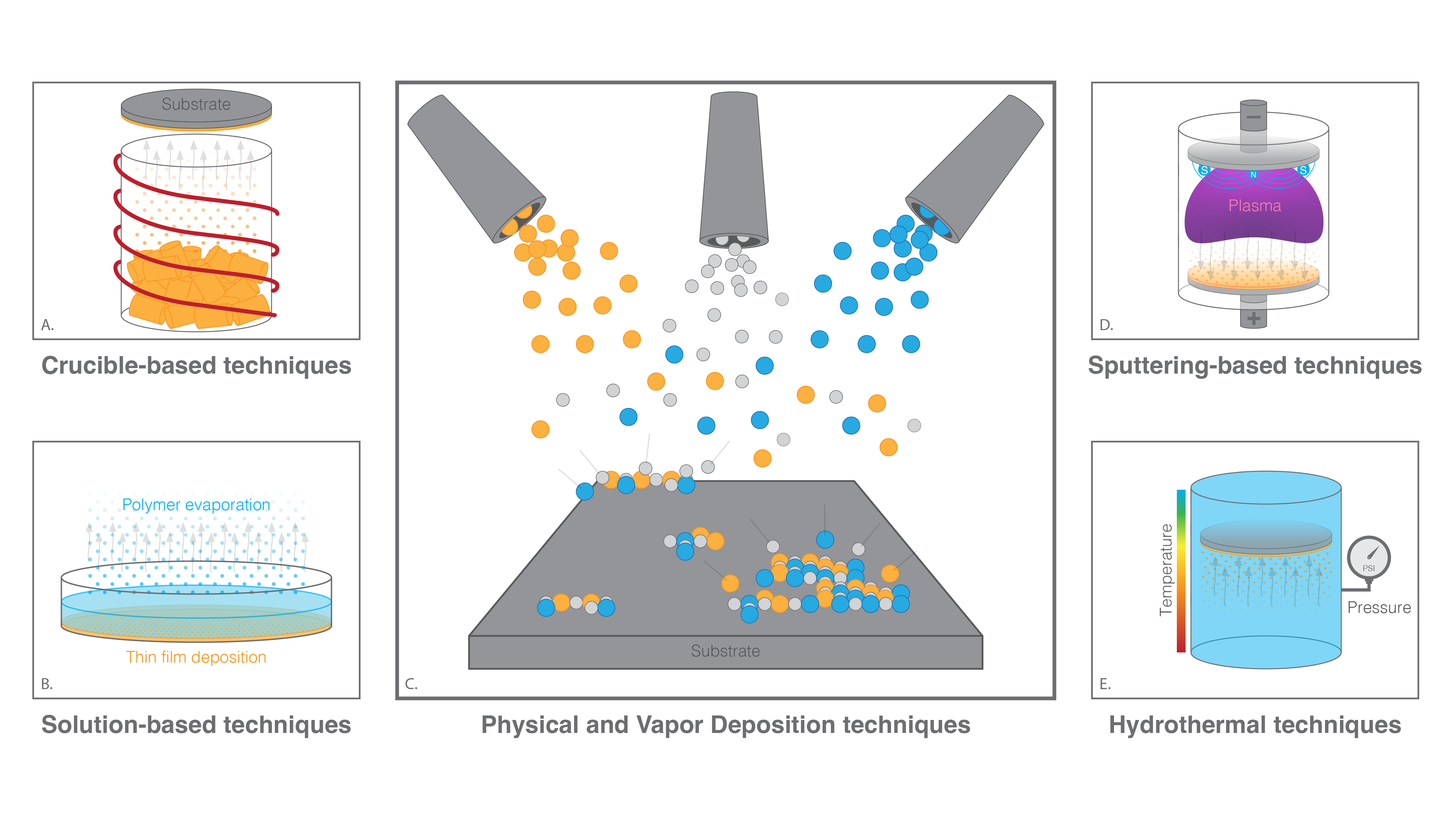}
    \caption{Schematic representation of the main synthesis techniques used for actinide thin film fabrication. (a) Shows a crucible often used in pyrometallurgical techniques where materials are evaporated from a solid source onto a substrate above. (b) Shows the polymer assisted deposition approach, where as a polymer solution evaporates, the film grows. (c) Depicts a common approach to vapor deposition techniques, where ultra pure material (either in elemental form or in compounds) is directed towards a substrate where it can react with a plasma, other compounds, or simply adsorb on the substrate. (d) Shows a representation of sputtering techniques, where a plasma of ionized material is deposited onto a substrate, with the option to have an overpressure of a different material, such as N or O. (e) Shows the hydrothermal approach, where pressure and temperature gradients are used to precipitate the desired material from a mineralized solution onto a substrate where a thin film can grow.}
    \label{fig:techniques}
\end{figure*}

Spirlet and Voigt summarized the first attempts at refining actinide metals and compounds~\cite{spirlet1982preparation}. Their review article includes the metallothermic reduction of oxides and halides, as well as vacuum melting, selective evaporation and condensation, electrorefining, zone melting, and electrotransport (see \cite{spirlet1982preparation} and references therein). They highlight solution-based techniques as the most economical approach for single crystal synthesis. In the 40 years since Spirlet and Voigt's review, growth of high quality single crystal actinides has advanced enormously, and these samples are of increasing interest from a fundamental scientific perspective. As safety standards evolve and more institutions have access to actinide sources, here we discuss the different approaches that will be instrumental in helping single-crystal actinides reach their full potential. Figure \ref{fig:techniques} summarizes several of these techniques including pyrometallurgical, solution-based, and different chemical, physical, and pulsed vapor deposition techniques. Table \ref{tab:mbe1} also lists these synthesis methods, together with the different compounds they have been used to produce, and substrates when available.

\subsection{Pyrometallurgical methods}

In early attempts to fabricate bulk single crystals, actinide compounds such as oxides or alloys were placed in crucibles. When subjected to extreme temperatures and pressures, the required materials are evaporated from the solid source onto a substrate (Figure \ref{fig:techniques}(a)). The high melting points of actinide and rare-earth materials used in this early attempts presented a challenge for substrate and crucible material selection. Among the first reports attempting to grow actinide thin films was \emph{mineralization}. In this approach, the crucible is placed on a rotating, water-cooled support, connected to a DC power supply. A tungsten wire acts as a cathode, emitting electrons that weld a lid onto the crucible. Then, by heating the crucible very close to its melting temperature, single crystals can grow inside from the source material over a period of weeks~\cite{mattenberger1986anisotropic,spirlet1982preparation,durakiewicz2004electronic}. In this way, Ac$_2$O$_3$ was chemically converted to a face-centered cubic (fcc) Ac metal powder, along with the presence of actinium hydride discovered through X-ray diffraction (XRD) ~\cite{farr1953crystal}. Similar approaches have yielded samples of Th$_3$N$_4$~\cite{bowman1971crystal}, UN~\cite{tagawa1974x}, U$_2$N$_3$~\cite{tagawa1974x}, USb~\cite{durakiewicz2004electronic}, UTe~\cite{durakiewicz2004electronic}, and U metal~\cite{radchenko2010production}. Different transuranic elements can be synthesized into thin films by reducing their oxide compounds. These materials are loaded into a Ta crucible, together with a reducing agent. The crucible is then heated with direct current filaments in order to evaporate the pure metal and condense it onto a cooler substrate outside of the crucible~\cite{radchenko2010production}. Pyrolitic decomposition has been used to thermally break down precursor molecules containing U and Am, and form thin films of these active species~\cite{danilin2002production}. In general, these early methods introduced large numbers of impurities, and resulted in poor control over the crystal quality of each sample. Nonetheless they helped researchers improve their understanding of growth parameters and physical properties, paving the way for more innovative approaches.

\subsection{Solution-based methods} 
The levels of radioactivity in some actinide materials make it impractical to handle of more than trace amounts in facilities not designed for this purpose. Dissolving these trace amounts in solution helps minimize the amount of radioactive elements involved in single-crystal growth (Figure \ref{fig:characterization}(b)). One of the most successful approaches to solution deposition to date is polymer assisted deposition (PAD), where precursor molecules are made by binding the metal ions of interest to a polymer. As this polymer solution evaporates, the metal film grows. This technique allows researchers to bypass the need for vacuum equipment by controlling the growth process through the viscosity and homogeneity of the metal-bearing precursor polymers~\cite{kruk2021growth,burrell2007controlling,scott2014high,mark2013optical}. Electrodeposition is another solution-based approach. Atoms of the required metal are suspended in solution and mixed with a metallic stirrer that acts as the anode, while a suitable metal plate onto which the film will grow is kept near the cathode~\cite{getoff1969molecular,muzzarelli1966preparation,shimojima1964electrodeposition}. This approach has provided an opportunity to study the oxidation states in uranium oxides ~\cite{burrell2007controlling}, and to grow thin films of uranium oxides on foils~\cite{qiu2001characterization}.

Other solution-based approaches include hydrothermal synthesis, in use since 1839 when the first barium and strontium carbonate crystals were formed~\cite{laudise2004hydrothermal}. Pressure and temperature gradients are used to precipitate the desired material from a high purity feedstock~(Figure \ref{fig:characterization}(e)). A mineralized solution (e.g., CsF in the case of UO$_2$) is often used to help transport the nutrient to the substrate where the thin film can crystallize and grow~\cite{young2016work,dugan2018electrical}. The main difference between conventional solution growth and hydrothermal synthesis is the viscosity of the liquid. Several actinide compounds have been successfully synthesized using solution approaches, including ThO$_2$~\cite{huentupil2019photochemical,rickert2022raman}, UN$_2$~\cite{scott2014high}, UO$_2$~\cite{rickert2021hydrothermal,burrell2007controlling,scott2014high,qiu2001characterization,young2016work,dugan2018electrical,majumdar2021irradiation}, U$_3$O$_8$~\cite{scott2014high}, UO$_3$~\cite{kruk2021growth}, UC$_2$~\cite{scott2014high}, NpO$_2$~\cite{scott2014high,mark2013optical}, and PuO$_2$~\cite{scott2014high}.

A key challenge associated with solution-based growth approaches is presence of trace amounts of solvent in the final samples. Although the quality of thin films prepared in the last few years is remarkable, further reductions in defect and impurity densities is still a goal for higher precision measurements.

\subsection{Physical and chemical vapor deposition methods} 
Several types of physical vapor deposition (PVD) have been used to synthesize thin films of actinides and surrogate materials (Figure \ref{fig:characterization}(c)). Among these, sputtering and electron beam evaporation (EBE) have been most commonly used due to their comparatively low cost and ease of use. Different types of sputtering are used for the deposition of thin films depending on the physical properties of the targets required: radio frequency (RF), direct current (DC), high frequency (HF), ion beam (IB), and magnetron sputtering are among the most common. Multiple types can be combined to produce enhanced results in terms of faster deposition rates or more uniform films (Figure \ref{fig:characterization}(d)). EBE is commonly used for its ability to heat relatively small areas of a source to very high temperatures, allowing for high precision in deposition rates. 

Reactive PVD (RPVD) occurs when a chemical reaction occurs between the deposited material and the atmosphere surrounding the substrate, often during an annealing step. In contrast, pulsed laser deposition (PLD) is a form of physical vapor deposition where a high power, low bandwidth laser is used to melt, evaporate, and ionize material from the surface of a target~\cite{enriquez2020structural}. 

Other techniques used mostly for rare-earth compounds include chemical vapor deposition (CVD)~\cite{straub2019chemical,raauf2021magnetic} and molecular beam epitaxy (MBE)~\cite{tarnawska2010single,ragnarsson2001molecular}. There have been numerous successful research efforts to use rare-earths as surrogates for actinide materials \cite{mori2019thermoelectric,ohkubo2021control,li2015gdn,goh2017lanthanide,leskela2006rare}. Similarities in their densities, electronic structure, and high melting temperatures give researchers an opportunity to develop equipment and growth regimes that can be translated from the rare earths to synthesize actinide thin films.

\section{Actinide Thin Films and Characterization}\label{sec:actinide}

Commercial applications of actinide-based compounds depend on a thorough understanding of their properties. This knowledge in turn helps researchers develop ways to synthesize higher quality samples. The accuracy of current physical property measurements for actinide-based materials are limited by defects such as grain boundaries and impurities in the samples. Initial studies of the properties of these materials used polycrystalline samples or, in the best cases, samples with large single crystal grains. Defects such as grain boundaries act as scattering centers for thermal energy carriers (electrons and phonons). Electron- and phonon-mediated conduction processes dictate a material's thermal and mechanical properties, and understanding these is vital for their full characterization. By eliminating grain boundaries, high quality, single crystalline samples enable researchers to study these processes in the absence of major scattering mechanisms. In addition, the use of new sample geometries and crystallographic orientations continue to expose novel physics and materials properties. For example, controlling dimensionality and strain in actinide heterojunctions will create an opportunity to explore novel quantum phenomena. Here we discuss the promise, challenges, and possible synthesis routes for actinide-bearing heterostructures exhibiting complex electron correlations for functional and energy materials. 

\subsection{Thin film synthesis} 
Thin films of actinium metal have been produced since 1953, while thin films of metallic protactinium were synthesized through electrodeposition as early as 1964~\cite{shimojima1964electrodeposition}. These metallic thin films were used for $\alpha$-spectrometry and nuclear reaction experiments. Following this work, deposition of thin films of protactinium pentoxide (Pa$_2$O$_5$) by electro-spraying films was achieved~\cite{muzzarelli1966preparation}. 

Reports of thorium-based thin film properties remain relatively scarce due to the absence of a 5$f$ electron orbital. Instead, most attention has focused on thorium's bulk properties as a candidate fuel for nuclear reactors. With this objective in mind, films of ThO$_2$ have been synthesized using sputtering~\cite{johnson2005thorium,brimhall2006thorium}, photochemical deposition~\cite{huentupil2019photochemical}, and electron beam evaporation~\cite{bagge2012morphology}. Uno \etal used a nitridation approach to synthesize Th$_3$N$_4$~\cite{uno1987preparation}. Gouder \etal prepared thin films of ThN and Th$_3$N$_4$ using sputter deposition, where photoemission spectroscopy demonstrated a high density of 6$d$ states in ThN and a non-metallic character in Th$_3$N$_4$~\cite{gouder2002synthesis}. 

UN thin films have been grown via reactive and DC magnetron sputtering using a Nb buffer layer on the $(1\bar{1}02)$ facet of sapphire~\cite{bright2018epitaxial}, as well as on Si(111), polycrystalline Ta foil and glass substrates~\cite{black2001preparation}. Black \etal used the same set of substrates (Si(111), polycrystalline Ta foil, and glass) in attempts to grow U$_2$N$_3$~\cite{black2001preparation}. Scott \etal synthesized UN$_2$ on LAO(001) substrates using polymer assisted deposition~\cite{scott2014high}. Long \etal also achieved the growth of both of these compounds by nitriding U metal flakes~\cite{long2015un2}. Gouder \etal grew this compound on CaF$_2$ using DC magnetron sputtering~\cite{gouder1997electronic}. Adamska \etal grew UMo films on a Nb(110) buffer layer on a sapphire substrate as well as U-Zr, U-Mo, and U-Nb~\cite{adamska2014growth}. The study by Black \etal also included thin films of UC$_2$ grown on YSZ(110) and U$_3$O$_8$ in different crystal structures that depended on the orientation of the sapphire substrates~\cite{black2001preparation}. Thin films of uranium oxicarbides and oxinitrides were prepared on Si(111) substrates by Eckle and Gouder using reactive DC sputtering~\cite{eckle2004photoemission}. Two compounds of uranium oxide were grown using PAD on LSAT(100) substrates: $\alpha$-UO$_3$ and $\alpha$-U$_3$O$_8$~\cite{kruk2021growth}.

Several attempts at synthesizing UO$_2$ thin films have been reported with various results. The first reported attempt is by Bierlein and Mastel in 1960, where they studied the effects of irradiation by growing UO$_2$ on carbon substrates~\cite{bierlein1960damage}. Elbakhshwan and Heuser summarized the evolution of UO$_2$ thin film growth up to 2017~\cite{elbakhshwan2017structural}, and discusses the use of CVD and sputtering techniques (see Table \ref{tab:mbe1}). Their contribution also includes growth on several different substrates using magnetron sputtering. Further advances in this field since then include reactive sputtering on LSAT(110)~\cite{teterin2016xps,popel2016structuralu}, magnetron sputtering on YSZ(100)~\cite{rennie2018study}, DC sputtering on SrTiO$_3$(110)~\cite{maslakov2018xps}, PAD of UO$_{2+x}$ by Zhang \etal~\cite{zhang2021optical}, and spin coat combustion on aluminum sheets by Roach \etal~\cite{roach2021hyperstoichiometric}.  U$_x$Th$_{1-x}$O$_2$ thin film synthesis was achieved by Cakir \etal to study surface reduction using ice~\cite{cakir2015surface}.

Synthesis of transuranic compound thin films has been historically challenging due to the high toxicity and scarcity of material precursors~\cite{miljevic2009environmental}. An early attempt to produce thin films of PuO$_2$ is reported by Shaw \etal~\cite{shaw1958preparation}. More recently, Wilkerson \etal summarized the state-of-the-art and reported the growth of thin films using PAD on YSZ substrates~\cite{wilkerson2020structural}. Roussel discussed the sputtering of PuO$_2$ and $\alpha$-Pu$_2$O$_3$ films used to study inverse photoemission~\cite{roussel2021inverse}. Subsequent attempts to grow bulk single crystals of these oxides used solution based approaches~\cite{finch1970high,finch1972high, modin2011indication}. Other Pu-based single crystal compounds synthesized include PuAs, PuSb, and PuBi for studies of anisotropic magnetization ~\cite{mattenberger1986anisotropic}. Mannix \etal used the same fabrication method to grow NpS and PuSb single crystals to investigate their magnetic structure~\cite{mannix1999experiments}. Durakiewicz \etal investigated the spectral features of USb, NpSb, PuSb, NpTe and PuTe using angle-resolved photoemission spectroscopy (ARPES) on samples grown using a mineralization technique~\cite{durakiewicz2004electronic}. Gouder \etal grew thin films of PuSe and PuSb for study using ultraviolet photoelectron spectroscopy (UPS)~\cite{gouder20005}, and thin films of PuSi$_x$ (with $x$ varying from 4 to 0.5)~\cite{gouder1997electronic}. A particularly successful research program involved PAD of PuO$_2$ and NpO$_2$ with high enough crystalline quality that they enabled studies of band dispersion and the Fermi surface, and helped validate several characteristics used in 5$f$ electron  modelling approaches~\cite{scott2014high}. NpO$_2$ was grown on LAO(001) substrates using a UO$_2$ buffer layer.  Elements beyond Np have scarcely been fabricated in sufficient amounts to attempt high quality crystal growth. Through a method of oxidation, Radchenko \etal managed to fabricate high quality bulk samples of Am, Cm, Bk, and Cf on a La crucible~\cite{radchenko2010production}.

\scriptsize
\begin{table*}[ht!]
\caption{Actinide thin film material growth summary.}
\label{tab:mbe1}
\begin{tabular}{|c|c|c|c|}
\hline
\textbf{Compound} & \textbf{Growth Method} & \textbf{Buffer Layer(s) - Substrate} & \textbf{Reference}\\
\hline
Ac/AcH$_2$ & Solid reaction & & \cite{farr1953crystal} \\
\hline
Pa & Solution & Zn, Al, Mn & \cite{morss2006chemistry} \\
\hline
ThO$_2$ & RF sputtering & various & \cite{johnson2005thorium} \\
        & RF sputtering & glass/Si & \cite{brimhall2006thorium} \\
        & Photochemical deposition & polyimide membrane & \cite{huentupil2019photochemical} \\
        & EBE & poly-Ir & \cite{bagge2012morphology} \\
        & Hydrothermal & & \cite{rickert2022raman} \\
        \hline
ThN     & metal reaction & N/A & \cite{bowman1971crystal} \\
        & DC sputtering & quartz & \cite{gouder2002synthesis} \\
        \hline
Th$_3$N$_4$ & Nitridation & N/A & \cite{uno1987preparation} \\
             & DC sputtering & Si(111) & \cite{gouder2002synthesis} \\
             \hline
UN & Reactive sputtering & Si(111)  & \cite{black2001preparation} \\
   &                     & poly-Ta foil & \\
   &                     & glass        &   \\
   & DC magnetron sputtering &
   $\langle 011 \rangle_{\text{Nb}} \| \langle1\bar{1}02\rangle_{\text{Al}_2\text{O}_3}$ & \cite{bright2018epitaxial} \\
   & Reactive sputtering & Spectrosil quartz glass & \cite{rafaja2005real} \\
   & DC sputtering & quartz glass & \cite{kim2010ion} \\
   & Cathode sputtering & Niobium & \cite{wang2016study} \\
   & Nitridation & & \cite{tagawa1974x}\\
   \hline
UN$_{1.85}$ & RF sputtering & Si & \cite{luo2020insight} \\
\hline
$\alpha$-U$_2$N$_3$ & Nitridation & & \cite{tagawa1974x} \\ 
\hline
    & Cathode sputtering & Si(100) & \cite{long2016study} \\
$\beta$-U$_2$N$_3$ & Nitridation &  & \cite{tagawa1974x} \\
\hline
U$_2$N$_3$ & Reactive sputtering & Si(111) & \cite{black2001preparation}\\
           &                     & poly-Ta foil &                       \\
           &                     & glass        &                        \\
    & DC magnetron sputtering & CaF$_2$ &\cite{bright2018epitaxial} \\
    \hline
UN$_2$  & PAD & $\langle111\rangle_{\text{UN}_2}\|\langle001\rangle_{\text{LAO}}$ & \cite{scott2014high} \\
\hline
UO$_2$& Reactive sputtering & LAO (001) / CaF$_2$ (100) &\cite{bao2013antiferromagnetism}\\
   & Reactive sputtering & $\langle001\rangle_{\text{UO}_2}\| \langle011\rangle_{\text{LSAT}}$ & \cite{popel2016structural,teterin2016xps} \\ 
   & Magnetron sputtering & $\langle001\rangle_{\text{UO}_2}\| \langle001\rangle_{\text{YSZ}}$ & \cite{popel2016structuralu} \\
   & Magnetron sputtering & $\langle111\rangle_{\text{UO}_2} \| \langle111\rangle_{\text{Si}}$ & \cite{chen2010structural} \\
   & Sputtering & Mo & \cite{senanayake2005coupling} \\
   & Chemical vapor deposition & UO$_2$ on Si & \cite{raauf2021magnetic} \\
   & Pulsed laser deposition & $\langle100\rangle_{\text{UO}_2}\| \langle111\rangle_{\text{LSAT}}$ & \cite{enriquez2020structural} \\
   & PAD & $\langle001\rangle_{\text{UO}_2} \| \langle011\rangle_{\text{LAO}}$ & \cite{burrell2007controlling, scott2014high} \\
   & sol-gel & {MgO, Al$_2$O$_3$} & \cite{meek2005some}\\
   & Solution/foil & Fe & \cite{qiu2001characterization} \\
   & Solution Spin Coat & Al & \cite{majumdar2021irradiation} \\
   & DC sputtering & $\langle001\rangle_{\text{UO}_2}\| \langle011\rangle_{\text{SrTiO}_3}$ & \cite{rennie2018study} \\
   & DC sputtering & $\langle111\rangle_{\text{UO}_2}\|\langle111\rangle_{\text{YSZ}}$ & \cite{maslakov2018xps} \\
   \hline
   \end{tabular}

\end{table*}
\normalsize
   
\scriptsize
\begin{table*}[ht!]
\caption{Actinide thin film material growth summary (cont.)}
\label{tab:mbe2}
\begin{tabular}{|c|c|c|c|}
\hline
\textbf{Compound} & \textbf{Growth Method} & \textbf{Buffer Layer(s) - Substrate} & \textbf{Reference}\\
\hline   
UO$_2$ & Hydrothermal & CaF$_2$ & \cite{young2016work,dugan2018electrical} \\
   & Hydrothermal & CaF$_2$,ThO$_2$,UO$_2$,YSZ & \cite{rickert2021hydrothermal} \\
   & CVD & Si & \cite{straub2019chemical} \\
   & Reactive Magnetron Sputtering & TiO$_2$, Al$_2$O$_3$, YSZ, ZnO, NdGaO$_3$ & \cite{elbakhshwan2017structural} \\
   & PAD & Si & \cite{zhang2021optical} \\
   & Spin coat combustion & Al sheet & \cite{roach2021hyperstoichiometric} \\
   \hline  
U-O-Pd & Sputter co-deposition & Si & \cite{stumpf2010development} \\
\hline  
U$_3$O$_8$ (P$\bar{6}$2m) & PAD & Al$_2$O$_3$(001) & \cite{scott2014high} \\
\hline  
U$_3$O$_8$ (Cmmm) & PAD & Al$_2$O$_3$(012) & \cite{scott2014high} \\
\hline  
$\alpha-$UO$_3$ & PAD & LSAT(100)  &\cite{kruk2021growth} \\
\hline  
$\alpha-$U$_3$O$_8$ & PAD & LSAT(100)  &\cite{kruk2021growth} \\
\hline  
UC$_x$O$_y$ & Reactive DC sputtering & Si(111) & \cite{eckle2004photoemission} \\
\hline  
UO$_x$N$_y$ & Reactive DC sputtering & Si(111) & \cite{eckle2004photoemission} \\
\hline  
U$_3$ & DC sputtering & Si(111) & \cite{gouder2004electronic} \\
\hline  
U$_{1-x}$Mo$_x$ & DC magnetron sputtering & Nb(110)/Al$_2$O$_3$(11$\bar{2}$0) & \cite{chaney2021tuneable} \\
\hline  
UMo & DC magnetron sputtering & Nb(110)/Al$_2$O$_3$(110) & \cite{adamska2014growth} \\
\hline  
USb & Mineralization & N/A & \cite{durakiewicz2004electronic} \\
\hline  
UTe & Mineralization & N/A & \cite{durakiewicz2004electronic} \\
\hline  
UC$_2$  & PAD &
$\langle001\rangle_{\text{UC}_2} \| \langle011\rangle_{\text{YSZ}}$ & \cite{scott2014high}\\
\hline  
U & Diode sputtering & poly-Al and Mg & \cite{gouder1997electronic} \\ 
    & Oxide reduction & La crucible & \cite{radchenko2010production} \\
    & UHV Magnetron sputtering & $\langle001\rangle_{\text{U}} \| \langle011\rangle_{\text{Nb/W}}\| \langle1\bar{1}02\rangle_{\text{\text{Al}$_2$\text{O$_3$}}}$ & \cite{ward2008structure,springell2014malleability}\\
    & Sputtering & Al and Mg & \cite{gouder1998thin} \\
    \hline  
UH$_2$ & Reactive sputtering & Si (001) & \cite{havela2018crystal} \\
\hline  
U$_x$Th$_{1-x}$O$_2$ & DC sputtering & Si(111) & \cite{cakir2015surface}\\
\hline  
(U$_x$Pu$_{1-x}$)O$_2$ & DC sputtering & Si(111) & \cite{cakir2015surface}\\
\hline  
NpH$_2$ & Solid-gas reaction & & \cite{mulford1965neptunium}\\
\hline  
NpSb & Mineralization & N/A & \cite{durakiewicz2004electronic} \\
\hline  
NpTe & Mineralization & N/A & \cite{durakiewicz2004electronic} \\
\hline  
NpO$_2$ & PAD & UO$_2$/LAO(001) & \cite{scott2014high} \\
& PAD & YSZ(100) & \cite{scott2014high,mark2013optical}\\
& DC sputtering & Si(111) & \cite{cakir2015surface}\\
\hline  
NpS & Mineralization & N/A & \cite{mannix1999experiments} \\
\hline  
PuN & Reactive Sputtering & Si(111) & \cite{havela2003photoelectron} \\
\hline  
PuH$_2$/PuD$_2$ & Solid-gas reaction & & \cite{mulford1955plutonium}\\
\hline  
PuSb & Mineralization & N/A & \cite{mattenberger1986anisotropic,mannix1999experiments} \\
\hline  
PuAs & Mineralization & N/A & \cite{mattenberger1986anisotropic} \\
\hline  
PuBi & Mineralization & N/A & \cite{mattenberger1986anisotropic} \\
\hline  
PuCoGa$_5$ & Flux growth & not specified & \cite{sarrao2002plutonium} \\
\hline  
PuSb & Mineralization & N/A & \cite{durakiewicz2004electronic} \\
\hline  
PuTe & Mineralization & N/A & \cite{durakiewicz2004electronic} \\
\hline  
\end{tabular}

\end{table*}
\normalsize

\scriptsize
\begin{table*}[ht!]
\caption{Actinide thin film material growth summary.}
\label{tab:mbe3}
\begin{tabular}{|c|c|c|c|}
\hline
\textbf{Compound} & \textbf{Growth Method} & \textbf{Buffer Layer(s) - Substrate} & \textbf{Reference}\\
\hline   
PuO$_2$ & DC sputtering & Si (111) & \cite{seibert2010interaction} \\
    & PAD & YSZ(100) & \cite{scott2014high,mark2013optical,wilkerson2020structural} \\
    & Sputtering & $\alpha$-Pu & \cite{roussel2021inverse} \\
\hline
$\alpha$-Pu$_2$O$_3$ & Sputtering & $\alpha$-Pu & \cite{roussel2021inverse} \\
\hline
PuSi$_x$ & DC sputtering & Mo (100) & \cite{gouder20005} \\
\hline
Am & DC Sputtering & Mo(100) & \cite{gouder2005photoemission}\\
\hline
AmN & DC Sputtering & Mo(100) & \cite{gouder2005photoemission}\\
\hline
Am$_2$O$_3$ & DC Sputtering & Mo(100) & \cite{gouder2005photoemission}\\
\hline
AmO$_2$ & DC Sputtering & Mo(100) & \cite{gouder2005photoemission}\\
\hline
AmSb & DC Sputtering & Mo(100) & \cite{gouder2005photoemission}\\
\hline
Am$_2$ & Solid-gas reaction &  & \cite{olson1966americium}\\
\hline
Am & Oxide reduction & La crucible & \cite{radchenko2010production} \\
\hline
CmH$_2$ & Solid-gas reaction & & \cite{BANSAL1970603}\\
\hline
Cm & Oxide reduction & La crucible & \cite{radchenko2010production} \\
\hline
Bk & Oxide reduction & La crucible & \cite{radchenko2010production} \\
\hline
Cf & Oxide reduction & La crucible & \cite{radchenko2010production} \\    
\hline
\end{tabular}

\end{table*}
\normalsize

\subsection{Characterization of Thin Films}
\begin{figure*}
    \centering
    \includegraphics[width=\textwidth]{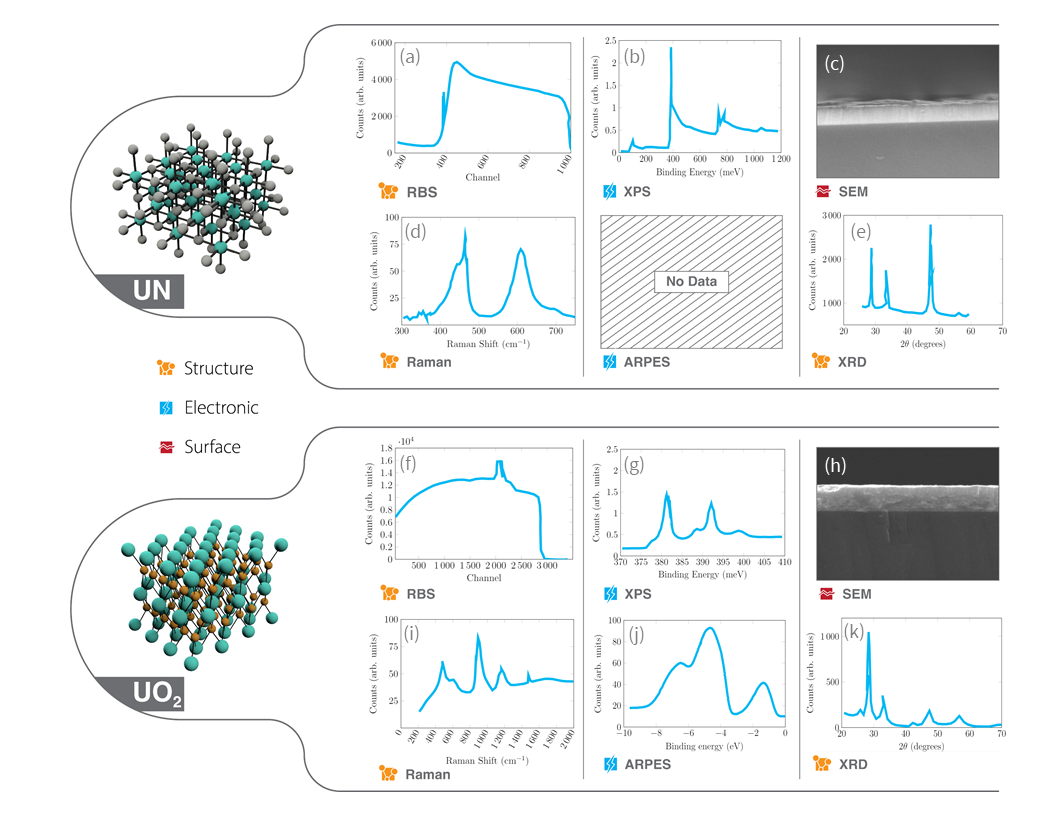}
    \caption{Characterization techniques applied to the most commonly studied actinide thin films: UN and UO$_2$. The data presented in each panel have been digitally extracted from the relevant references and are shown for illustrative purposes only. (a) RBS data from UN before and after radiation exposure (\cite{kim2010ion}), (b) XPS data for a UN thin film (~\cite{bright2018epitaxial}), (c) SEM micrograph of a UN thin film (~\cite{wang2016study}), (d) UN Raman spectrum \cite{luo2020insight}, and (e) an XRD diffraction pattern for UN \cite{lu2018thermal}. The equivalent data for UO$_2$ come from (f) \cite{apostol2017ion}, (g) and (h) \cite{chen2010structural}, (i) \cite{enriquez2020structural}, (j) \cite{scott2014high}, and (k) from \cite{chen2010structural}. Note that ARPES data do not yet exist for UN thin films, making initial electronic characterization a near-term target for any new synthesis.}
    \label{fig:characterization}
\end{figure*}

We can divide materials characterization techniques into categories that deal with different properties of the thin films in question. Figure \ref{fig:characterization} illustrates how one can use various techniques to explore structural, surface and electronic properties together in concert.

We can obtain plan view or cross-sectional structural images of a thin film using transmission electron microscopy (TEM), scanning electron microscopy (SEM) (see Figure \ref{fig:characterization}(c) and (h)), atomic force microscopy (AFM), and scanning tunneling microscopy (STM). These can be used to determine the size of crystallites, the density of defects such as dislocations twins and stacking faults, and crystalline quality, in addition to quantifying the roughness of the surfaces and the shape, size and areal density of any structures present on them. ``Indirect" structural characterization may include Rutherford back scattering (RBS), XRD, and Raman spectroscopy (see Figure \ref{fig:characterization}(a), (d), (e), (f), (i), and (k)). These  techniques can provide information that complements what can be seen in microscopy, and help identify properties such as alloy composition, lattice parameter, and in the case of ion-damaged samples, penetration depth.

Several techniques have been used to measure the thermophysical properties of bulk actinide materials~\cite{hurley2021thermal}. However, thin films have been targeted in specific cases as a means to increase measurement accuracy while using non-destructive techniques. It is here where photothermal approaches offer a viable option for studying samples with small physical dimensions without damage to the sample or the need for destructive sample preparation approaches~\cite{dennett2020influence}. Several comprehensive reviews have been recently published on the thermophysical properties of actinide oxides~\cite{hurley2021thermal}, nitrides~\cite{parker2021thermophysical}, and carbides~\cite{jackson2010laser,daroca2013phonon} due to the importance of these compounds to nuclear science and industry.

Electronic structure characterization includes angle-resolved photemission spectroscopy (ARPES), X-ray photoemission spectroscopy (XPS), STM-based scanning tunneling spectroscopy (STS), and photoluminescence spectroscopy. These approaches can shed light 
on charge carrier behavior, and the electronic states present in a given material. These characteristics are of particular interest for actinide thin films due to the interesting physics governed by their 5$f$ electron states. XPS allows one to determine the elements present within a material, along with their oxidation and bonding state. XPS is especially useful for samples with reduced dimensionality within which the localization of 5\textit{f} electrons occurs~\cite{gouder2002examples}. Gouder and Havela identified the complications of applying XPS to the study of actinides: an asymmetrical tail in metallic systems can often prove difficult to differentiate from scattering of electrons onto the Fermi level. In addition, the anisotropic distribution of atoms at different temperatures contributes to higher noise levels in the XPS signal~\cite{gouder2002examples}. The team lead by Black \etal  identified the 5$f$ states of UN around 0.8 eV \cite{black2001preparation}. Havela \etal indicate that 5$f$ localization behavior can be studied in PuSb (localized), PuSe (intermediately localized), and PuN (potentially delocalized)~\cite{havela2003photoelectron}.  In addition, Gouder \etal studied the localization of 5\textit{f} electrons in PuSi$_x$~\cite{gouder2004electronic}. Durakiewicz \etal investigated the systematic changes in the 5\textit{f} binding energy peak between antimonide and tellurides of U, Pu, and Np~\cite{durakiewicz2004electronic}. Joyce \etal studied the dual nature of 5\textit{f} electrons using photoemission of $\delta$-Pu, PuIn$_3$, and PuCoGa$_5$~\cite{joyce2006dual,joyce2011pu}. On the same line, Eloirdi \etal reported results from XPS studies of PuCoGa$_5$ deposited using sputtering~\cite{eloirdi2009photoelectron}. Long \etal studied the electronic structure of $\alpha$-U$_2$N$_3$ and found hints of metallic bonding~\cite{long2016study}. Pentavalent uranium was found in U$_2$O$_5$ by Gouder \etal using photoemission studies~\cite{gouder2018direct}.

The ARPES technique has been used to determine precise band structures and Fermi surfaces of actinide-based single crystals~\cite{scott2014high,sarrao2002plutonium,fujimori2012itinerant,durakiewicz2008observation,das2012imaging,kawasaki2013band,beaux2011electronic,denlinger2001comparative,ito1999band,santander2009fermi,yoshida2010signature,kawasaki2011electronic,dakovski2011anomalous,yoshida2012observation,kawasaki2011band,boariu2013momentum,chatterjee2013formation,yoshida2013translational,meng2013imaging,bareille2014momentum,durakiewicz2014photoemission,ruello2004thermal,tobin2011orbital,gilbertson2014ultrafast,roy2008dispersion,joyce1998resonance}. Largely focused on bulk actinide compounds, these ARPES studies have created new opportunities for determining the origin of the unusual physical properties of 5$f$-electron materials. Going forward, ARPES-based studies of epitaxial actinide thin films could lead to the discovery of a wide range of emerging electronic, magnetic, and structural phenomena (Figure \ref{fig:characterization}(j)). Interactions between the epitaxial thin film and the underlying substrate can produce properties that are substantially different from those of the bulk materials. For instance, it has been reported that one can obtain an unusual hexagonal close-packed phase of uranium (hcp-U) by depositing U thin films onto a W (110) substrate~\cite{molodtsov1998dispersion,berbil2004observation}. Theoretical calculations predict that the hcp-U phase has an electronic instability, that could lead to a possible charge density wave or magnetic ordering~\cite{zarshenas2012theoretical}. Different orientations of the $\alpha$-U orthorhombic phase can also be obtained by depositing U onto a variety of buffer/seed layers on sapphire substrates~\cite{ward2008structure,springell2008elemental}. ARPES was used to obtain the band structure of $\alpha$-U single crystals at 173~K~\cite{opeil2007angle}. The valence band structure of $\alpha$-U was studied in details using ultraviolet photoemission spectroscopy and {X}-ray photoemission spectroscopy~\cite{opeil2006valence}, 

For ordered overlayers of U metal grown on a W (110) substrate, band-like properties of the U 5$f$ states were observed, which were proposed to arise from direct $f$–$f$ electron interactions~\cite{molodtsov1998dispersion}. STM/STS results showed that the U density of states close to the Fermi level is dominated by the 5$f$ states~\cite{berbil2004observation}. STS and ARPES showed both the Fermi surface topology and electronic structure of ordered $\alpha$-U films~\cite{chen2019direct}. ARPES studies of epitaxial thin films of the oxides PuO$_2$ and NpO$_2$~\cite{mark2013optical,joyce20105f}, and uranium carbide (UC$_2$), reveal the importance of the underlying substrate structure in stabilizing the epitaxial film~\cite{jilek2013preparation}. But in general, compared with bulk actinide crystals, only a few ARPES studies have been performed on actinide thin films. Therefore, the use of ARPES to measure novel actinide thin films synthesized with higher quality will produce results that are critical to understanding their reactivity, aging, nuclear fuel behavior, and environmental fate, as well as the potentially rich physics of these exotic 5$f$ materials.

\section{Thermophysical, Magnetic, and Topological Properties of Actinide Thin Films}\label{sec:properties}

\subsection{Thermophysical Properties}
Thermophysical properties play a central role in determining the characteristics of actinide materials. Atomic structure dictates macroscopic behavior, and it is hence important to understand the temperature dependence of these properties. A key research driver in actinide compounds is the qualification process for nuclear fuel materials. This process is heavily reliant on characterizing thermophysical properties: heat capacity, thermal conductivity, elastic constants, and thermal expansion~\cite{hurley2021thermal,middlemas2020determining}. Heat capacity has been measured in several important actinide dioxides, such as ThO$_2$~\cite{dennett2020influence}, UO$_2$~\cite{sanati2011elastic}, NpO$_2$~\cite{santini2009multipolar}, PuO$_2$~\cite{magnani2005perturbative}. UO$_2$ and NpO$_2$ show an interesting magnetic alignment transition that creates anomalies in their heat capacity. Thermal conductivities have been measured for ThO$_2$~\cite{mann2010hydrothermal}, UO$_2$~\cite{gofryk2014anisotropic}, NpO$_2$~\cite{nishi2008thermal}, and PuO$_2$~\cite{lagedrost1968thermal}. The heat capacities of select actinide nitride systems have been measured, including ThN~\cite{parker2019thermophysical}, and more recently for the UN-ThN mixtures used in nuclear fuels~\cite{parker2021thermophysical}.

Among these properties, thermal diffusivity and conductivity have been recently studied as important factors in determining the properties of a nuclear fuel candidate material~\cite{hurley2021thermal}. The qualification process requires that intrinsic thermophysical properties be resolved independently of any particular specimen geometry, without the competing influence of lattice defects or other chemical or structural imperfections. Building a fundamental understanding of intrinsic properties allows for the effects of these additional complexities, for example the types of structural defects generated under exposure to irradiation, to be systematically included in computational frameworks for engineering-scale performance. Much success to date has been found in the use of non-destructive photothermal techniques to measure these intrinsic thermophysical properties~\cite{dennett2020influence}. Such methods broadly consider microscale thermal and acoustic wave propagation, which can be measured via surface deformation, refractive index gradient, acoustic expansion/contraction, or changes in optical reflectivity (thermoreflectance). This last approach can be further subdivided into frequency-domain thermoreflectance, time-domain thermoreflectance, spatial-domain thermoreflectance, and hybrid methods~\cite{middlemas2020determining,khafizov2017investigation}. With small sampling volumes, these methods are broadly applicable to the study of thin films~\cite{luckyanova2013}. Although they have not yet been applied to actinide thin films, we anticipate that these techniques will help to shed light 
on their thermophysical properties in the future.

Heterostructure devices that integrate different functional layers can encounter performance issues as a result of restricted heat dissipation across the heterointerfaces (Kapitza resistance). This restriction in heat dissipation correlates with factors such as interfacial roughness, disorder, dislocations, and bonding. As a result, computational modelling of these interfaces often fails to capture all of their vibrational properties, for example changes in mass density, phonon mean free path, and carrier scattering across the interface~\cite{giri2016effect}.

\begin{figure*}
    \centering
    \includegraphics[width=\textwidth]{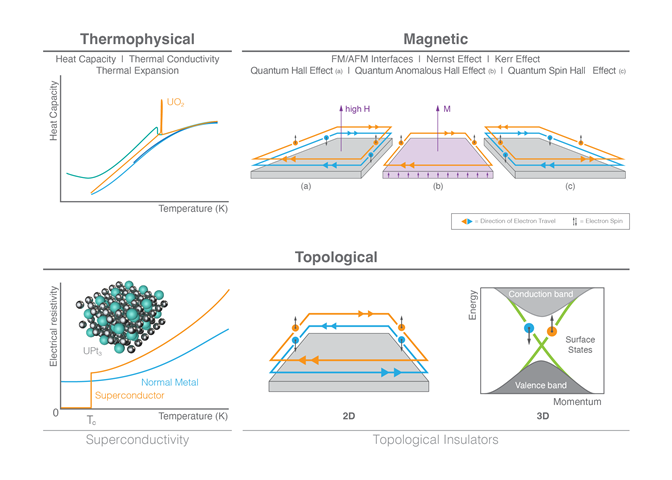}
    \caption{Properties of interest in select actinide thin films considered in this review. Under thermophysical properties, measurements of heat capacity in some actinide oxides exhibit an anomaly at certain phase transition temperatures. Magnetic properties to consider include the quantum, quantum anomalous, and quantum spin Hall effects. Two important topological properties are superconductivity (present in the UPt$_3$ compound, and topologically protected states) together with topological insulators in two and three dimensions, respectively.}
    \label{fig:properties}
\end{figure*}

\subsection{Magnetic properties}

Different aspects of electron and phonon behavior have been carefully studied under the influence of magnetic fields. Among these, transport and its different manifestations enabled the observations of several noteworthy phenomena. These physical properties are part of what makes actinides interesting to researchers beyond their nuclear applications~\cite{santini1999magnetism}. Most transition metal elements have itinerant $d$ electrons, while lanthanide elements exhibit mostly localized 4\textit{f} electrons. These fact help us distinguish characteristics related to the conduction band of a material and its magnetic properties relating to the 4\textit{f} electrons~\cite{santini1999magnetism}. Due to the scarcity and high cost of $^{227}$Ac, single crystal samples of this metal have been synthesized to date, making it difficult to study its magnetic properties experimentally~\cite{deblonde2021coordination}. A similar situation exists for the actinide elements after Californium. 

Considering the intervening elements, Th is a paramagnetic metal with three phases that are heavily influenced by the concentration of impurities present \cite{morss2006chemistry}. Pa metal was synthesized relatively early on in the development of actinide research, and it was hence possible to determine its phase transition and superconducting behavior \cite{morss2006chemistry}, as discussed below. The early actinide metals (U, Np, Pu) show no sign of magnetic ordering mostly because the bandwidth of the 5$f$ electrons in these elements is too wide to satisfy the Stoner criterion \cite{severin1993relationship}. U metal exhibits a weak Pauli-type paramagnetism, with $\alpha$-U having a charge-density-wave anomaly around 43 K that has been observed in magnetic susceptibility measurements \cite{morss2006chemistry}. Np exhibits spin-orbit coupling in the same order as other crystal-related electric effects, leading to its metallic form and paramagnetic behavior \cite{morss2006chemistry}. Elemental Pu can be found in both $\alpha$- and $\delta$-Pu phases. Recent studies point to a lack of evidence favoring any ordered or disordered magnetism in $\delta$-Pu \cite{lashley2005absence}, which preferentially forms a valence fluctuation ground state \cite{janoschek2015valence}. Curium (Cm) is the first element in the actinide series that orders magnetically. Cm is antiferromagnetic (AF) below 65 K in the double hexagonal close packed (dhcp) phase, while in the fcc phase it is ferromagnetic (FM)  below 205 K \cite{lander2019measurements}. Berkelium (Bk) also exists in these two crystallographic phases (dhcp and fcc) and similarly orders either AF or FM. Californium (Cf) also exhibits a large effective magnetic moment, similar to Cm and Bk, and likely has a FM ground state \cite{ott1987handbook}.

A variety of thin films of actinide compounds have provided insight into the magnetic properties of these systems, such as the magnetic structures of NpS and PuSb \cite{mannix1999experiments}, itinerant antiferromagnetism \cite{rafaja2005real} and  magneto-optical Kerr effect \cite{marutzky2006optical} in UN , weak ferromagnetism \cite{sakai2004magnetism} and antiferromagnetism \cite{bao2013antiferromagnetism} of UO$_2$, and the high pressure behavior of electrical resistance in NpAs and NpBi \cite{ichas1997electrical}.
Magnetic susceptibility measurements of transplutonium elements Am, Cm, and Bk helped determine the effective magnetic moments and Curie-Weiss constants for some of their compounds~\cite{nave1983magnetic}. Magnetic properties have also been extensively studied in these materials using computational methods (\cite{jin2018magnetic} and references therein). Bright \etal studied the magnetic and electronic structure of UN and U$_2$N$_3$ epitaxial films using x-ray scattering~\cite{bright2019synchrotron}. Recently, using high-field magnetization and magnetostriction measurements, the presence of a metamagnetic transition and a tri-critical point was observed in UN single crystals, leading to the development of a magnetic phase diagram for this material \cite{shrestha2017tricritical,troc2016electronic}. Using micro-structured single crystals, Hamann \etal found strong reconstruction of the Fermi surface at the high-field transition \cite{hamann2021fermi}.

\subsection{Superconductivity}
Certain actinide metals are known to exhibit superconductivity at low temperature. these include Th at 1.4 K \cite{wolcott1958superconductivity, fowler1965superconductivity}, Pa at 1.4 K \cite{smith1979superconducting}, $\alpha$-U at 0.7 K and $\gamma$-U at 1.8 K \cite{aschermann1942elektrische, goodman1950superconductivity, gordon1966superconductivity, moore2009nature}. Am was predicted to be a superconductor, although confirming this experimentally was a challenge due to its large self-heating, scarcity, and radioactivity. However, this prediction was confirmed by Smith and Haire; when its fcc phase is stabilized by quenching, Am becomes superconducting at 1 K  \cite{smith1978superconductivity}. $\alpha$-Np and $\alpha$-Pu have no reported superconducting signature down to 0.5 K \cite{griveau2014superconductivity} and Cm, Bk, and Cf are not expected to show superconductivity because of their magnetic ground-state~\cite{griveau2014superconductivity}. There are several actinide compounds that show heavy-fermion superconductivity, including UBe$_{13}$, UPt$_3$, URu$_2$Si$_2$, UNi$_2$Al$_3$, UPd$_2$Al$_3$, NpPd$_{5}$Al$_{2}$, PuCoIn$_{5}$, PuRhGa$_{5}$, PuRhIn$_{5}$ and PuCoGa$_{5}$~\cite{sarrao2002plutonium,maple1997superconductivity, aoki2007unconventional, griveau2008transport, bauer2011localized, wastin2003advances, bauer2015plutonium}. Ott \etal showed that UB$_{13}$ is superconducting below 0.85~K~\cite{ott1983u}. Joyce \etal studied photoemission data from PuCoGa$_5$, PuIn$_3$, and $\delta$-Pu metal. A comparison between their results and model studies indicate characteristics of both itinerant and localized Pu 5$f$ electrons~\cite{joyce2006dual}. Furthermore, Joyce compared $T_c$ in PuCoGa$_5$ against UCoGa$_5$ and found no evident characteristics of strongly correlated electrons~\cite{joyce2011pu}. Thus, valence fluctuations have been proposed as an alternative mechanism for the high-temperature superconductivity observed in PuCoGa$_5$~\cite{ramshaw2015avoided, gofryk2016thermoelectric}. Recently, unconventional \cite{ran2019nearly}, spin-triplet~\cite{nakamine2019superconducting} superconductivity with multiple superconducting phases \cite{braithwaite2019multiple} has been reported in UTe$_2$. In addition, there are indications for spontaneously broken time-reversal symmetry \cite{hayes2020weyl}, and chiral Majorana edge and surface states \cite{bae2021anomalous}, the nature of which are not yet understood. Brun, Cren, and Roditchev summarized the superconductivity of 2D materials as a strong case for epitaxial monolayer materials growth~\cite{brun2016review}.

\subsection{Topological Properties}

The study of conducting states in topologically nontrivial materials is often hindered by the presence of bulk conducting states~\cite{bhardwaj2020thickness}. High quality thin films offer one way of minimizing bulk contributions by increasing a sample's surface area-to-volume ratio, simplifying measurements of these topological effects. Due the nature of their electronic structure, actinide materials represent a fascinating, albeit challenging area for topological discoveries. Promising technological applications and insights into fundamental physics continue to motivate investigations on the complex manifold of 5$f$, 6$p$, 6$d$, and 7$s$ orbital shells~\cite{cossard2021charge}. These interactions, present in an analogous manner in the rare earths, are expected to give rise to a variety of topological behaviors in different compounds and geometries.

Heavy fermion materials owe their particular properties to the competition between Coulomb repulsion of localized $f$ electrons, and hybridization with itinerant $d$ electrons. These electron correlations produce complex states including Mott physics, superconductivity, and correlated magnetism \cite{richmond1968electron}. Understanding these electronic systems could potentially help elucidate other strongly correlated electron materials, for example high-temperature superconductors, multiferroics, and magnetoresistive materials. Topological insulators possess conducting surface states protected by the non-trivial topological property of spin-momentum locking. Several quantum materials have been proposed based on topological properties that involve quantum phenomena such as Majorana fermions, the anomalous quantum Hall effect, with applications in spintronics and fault tolerant quantum computation. Magnetic topological semimetals have been synthesized using MBE \cite{krumrain2011mbe,tang2017quantum}. These materials have a FM ground state that breaks time-reversal symmetry with a gap in the topological surface states~\cite{hesjedal2019rare}.
A powerful tool to study these phenomena is density functional theory (DFT), e.g. its use in identifying phases in UNiSn corresponding to a topological insulator and a Weyl semimetal depending on magnetic ordering~\cite{ivanov2019topological}. Rocksalt compounds of Pu and Am were predicted to form a topologically insulating ground state due to their 5$f$ electron being right on the boundary between localized and itinerant, and coupling between electronic and spin-orbit interactions gives rise to band gap formation~\cite{zhang2012actinide}. The predicted compounds that Zhang \etal used for this study include americium monopnictides and plutonium monochalcogenide~\cite{zhang2012actinide}. Using nuclear magnetic resonance (NMR), Dioguadi \etal demonstrated that PuB$_4$ is a strong candidate for a topological insulator, which has also been suggested by Choi \etal~\cite{choi2018experimental,dioguardi2019pu}. Recent theoretical work suggests that the intermediate valence PuB$_6$ is a strong topological insulator with nontrivial $\mathbf{Z}_{2}$ topological invariants \cite{deng2013plutonium}, similar to SmB$_6$ \cite{dzero2010topological}.

Overall, much work is still to be done in understanding fully the properties of these actinide compounds, and high quality thin films promise to be the missing piece to provide a window into their discovery.

\section{Challenges with Modelling 5\textit{f} Electrons}\label{sec:modelling}

DFT~\cite{Kohn19651133,Hohenberg1964864} is normally the zeroth level of first-principles theory that is applied to any material, given the relatively small computational cost for the degree of complexity it can capture~\cite{Burke2012150901,Jones2015897}. DFT has found success across the periodic table for calculating a myriad of materials properties~\cite{Cramer200910757,Hasnip201420130270,Jain201615004,Maurer20191}. Unfortunately, all known implementations of DFT have well-known limitations~\cite{Perdew20011,Kotliar2006865,Cohen2008792}. In particular, DFT sometimes \emph{qualitatively} breaks down for certain properties in materials bearing strongly correlated electrons. For example, Mott insulators are often incorrectly predicted to be metals by DFT.~\cite{Imada19981039} Materials with $f$-electrons are ripe for strongly correlated electron physics, due to the relatively narrow bands formed by the $f$-orbitals and large on-site Coulomb repulsion. Other limitations for DFT calculations, particularly in strongly correlated systems, include the inability to properly capture finite temperatures and excited states~\cite{Perdew20172801}.

Fortunately, dynamical mean-field theory (DMFT)~\cite{Georges199613} can overcome many of the shortcomings of DFT. DMFT can also be applied to real materials when combined with DFT to yield the DFT+DMFT formalism~\cite{Kotliar2006865}. DMFT requires the solution of a quantum impurity problem, which is associated with the correlated shell of electrons (e.g. $f$-electrons). These solutions can be achieved in realistic multiorbital problems using continuous time quantum Monte-Carlo (CTQMC)~\cite{Werner2006076405,Haule2007155113,Gull2011349}. While DFT+DMFT(CTQMC) is far more computationally expensive than DFT, current algorithms and computing power allow for calculations in relatively complex systems bearing $d$ and $f$ electrons~\cite{Marianetti2008056403,Park2014245133,Isaacs2020045146,Adler2019012504,Chakrabarti2017064403,Lechermann2020041002}. An efficient, but far less accurate solution of the DMFT impurity problem can be achieved with the Hartree-Fock method. This method yields the better known DFT+$U$ approach~\cite{Anisimov1997767,Kotliar2006865}, for which the computational overhead is negligible when compared to DFT.  Together, DFT, DFT+DMFT, and DFT+$U$ form a hierarchy of first-principles approaches that serve as the basis for parameterizing multiscale models to generate information at longer length and time scales~\cite{sahoo2020structural}. 

\begin{figure*}
    \centering
    \includegraphics[width=\textwidth]{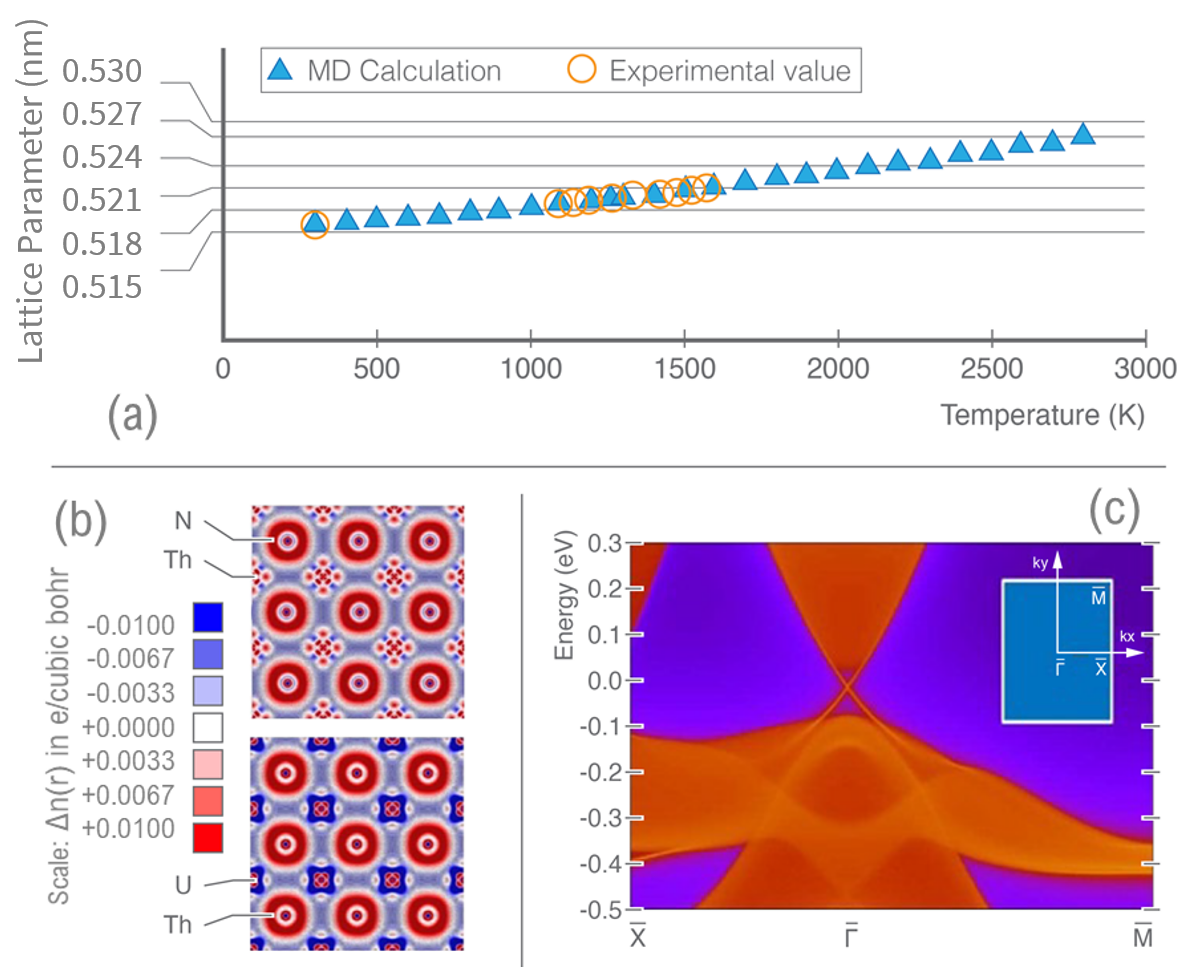}
    \caption{(a) Comparison between the experimental thorium nitride lattice parameter as a function of temperature, and values calculated using molecular dynamics (MD) simulations~\cite{adachi2005molecular}. Reproduced with permission from Elsevier. (b) Electron density plots calculated on the (001) plane for UN and ThN~\cite{atta2007density}. Reproduced with permission from the American Physical Society. (c) Band structure and energy spectrum of AmN, showing spin-orbit coupling interactions and a Dirac surface cone. From \cite{zhang2012actinide}. Reproduced with permission of AAAS. }
    \label{fig:modelling}
\end{figure*}

Compared with bulk materials, the surfaces and heterostructures associated with thin films introduce additional complexity to modeling efforts due to confinement effects, charge transfer, strain, and structural reconstruction. When considering actinide thin films, the presence of $f$-electrons in these systems contributes yet another layer of complexity. In actinide oxides, a wide range of phenomena have been observed in thin film heterostructures that are absent from their bulk constituents. These phenomena include superconductivity~\cite{Reyren20071196,castro2015147001}, room temperature ferromagnetism~\cite{Luders2009241102}, metal-insulator transitions~\cite{Boris2011937}, and orbital reconstruction~\cite{Chakhalian20071114}, with the clear potential for novel device physics~\cite{Bibes2011931912870}.  DFT+$U$ and DFT+DMFT have already been used to confirm emergent properties in heterostructures composed from transition metal oxides~\cite{Chen2013116403,Zhong2015246401,James2021023149}. However, similar calculations for heterostructures built from $f$-electron materials will undoubtedly be more challenging than those for $d$-electron materials. The DFT+DMFT computations will be substantially more computationally expensive, and DFT+$U$ computations will struggle with a more treacherous energy landscape.  Nonetheless, $f$-electron heterostructures are within reach of our present theories and computational resources, and there has already been some progress at the level of DFT.  Gao and Asok used DFT to study the (001),~\cite{gao2006convergence} and (111)~\cite{gao2008relativistic} surfaces of Am. The surface properties of UN, a candidate nuclear fuel, have been studied using DFT+$U$, with potential applications for understanding corrosion characteristics~\cite{sikorski2021first}. In short, a combination of models based on DFT, DFT+$U$, and DFT+DMFT are well poised to study heterostructures composed from $f$-electron material thin films, and will lay the foundation for understanding such materials. 

Optimization of algorithms based on strongly correlated electron systems in DFT will provide more accurate descriptions of physical systems, able to in turn provide higher level modelling systems a path towards integration with engineering-level modelling paradigms based on first principles.

\section{Future Directions: Thin film heterostructure synthesis and challenges}\label{sec:future}

Interfaces in semiconducting, metallic, and semi-metallic materials offer possibilities for investigating electronic properties not typically present in the bulk, or in layers of a single material. In non-actinide semiconductors, heterostructures built by stacking different thin films have enabled researchers to discover new physics. For example, thin films and other low dimensional heterostructures allow to study the quantum behavior of charge carriers whose de Broglie wavelength is equivalent to the size of the potential well within which they are confined. A deeper understanding of how we can manipulate charge carrier behavior using heterostructures led to the design of entirely new electronic and photonic devices. Indeed, Kroemer famously stated that in these heterostructures, the interface \textit{is} the device~\cite{kroemer2001quasi}. Extending heterostructure growth to actinide-based compounds therefore promises to enable researchers to study these materials in new ways. For example, low-dimensional heterostructures composed of actinide materials with 5$f$ electrons, grown with interfaces containing tailored defect densities, represent a fertile area for the discovery of various emerging phenomena~\cite{dennett2022towards}. Studies have predicted a range of exotic physics for such structures, including highly correlated electron interactions~\cite{cossard2021charge,block2021recent}, topological insulator behavior~\cite{dioguardi2019pu,ivanov2019topological}, and confined 2D properties~\cite{grover2020adsorption,lopez2020f}.

As we have seen, the synthesis of high quality, atomically precise thin films of a \textit{single} actinide material is still the focus of ongoing research. As a consequence, the ability to routinely and controllably integrate two or more of these thin films to create actinide heterostructures remains largely an area for future development. Before one can begin to study the characteristics of these heterostructures, one must first be able to produce high purity films with uncontaminated interfaces. Controlling interactions at the interface between two films requires a consideration of atomic intermixing, surface energies, charge transfer, and epitaxial nucleation and growth. These effects will likely be a function of the environment within which a heterostructure is synthesized~\cite{park2011interface}. Solution- and chemical reaction-based synthesis approaches, for example, often leave residues of unwanted atomic species during deposition. For this reason, physical deposition methods offer perhaps the best route to achieving pristine thin film formation, and hence heterostructure synthesis. Due to its rapid output and relatively low cost, pulsed laser deposition offers a convenient test-bed for initial film and substrate screening studies. After some of the film-substrate combinations predicted by theory have been demonstrated experimentally, even higher quality films can be grown using molecular beam epitaxy (MBE), which uses ultrahigh purity elemental source materials rather than chemical precursors or solutions. The advantages of MBE include exceptionally high material purity, and exquisite control over layer thickness and interfacial abruptness, both of which are prerequisites for growing functional heterostructures. To our knowledge, MBE has not yet been used for the synthesis of actinide thin films or heterostructures. 

The epitaxial challenges for integrating different materials include lattice constant mismatch, dissimilar crystal structures, thermal expansion differences, and incompatible growth environments. Several strategies have been devised to overcome strain-related issues arising from lattice mismatch. These include the use of graded buffers~\cite{murali2020study,Simon2011a}, interfacial misfit arrays~\cite{benyahia2018optimization}, self-assembled quantum dots~\cite{sautter2020strain}, alloying~\cite{makimoto2015new}, lattice constant mismatch inversion \cite{vallejo2021tensile}, and strain-relieving superlattices~\cite{he2021step}. To accommodate different crystal structures, a common crystal lattice parameter can be used to geometrically align the lattices between each layer. An example is the differing crystal structures of ErAs (rocksalt) and InGaAs (zincblende). A rotation of the ErAs unit cell with respect to that of the InGaAs brings the two lattices into registry, enabling the growth of coherent ErAs/InGaAs heterostructures~\cite{nandi2021material}. Mismatch between thermal expansion coefficients in different layers of the heterostructure can induce strain upon sample cooling, which is then relaxed through defect formation, negatively impacting crystal quality. Common solutions to this challenge include growth of thin protective thermal layers that prevent these relaxation processes, and use of thermally pre-stressed films that can accommodate the strain upon the change in temperature \cite{jin2005growth}. It will be critical to consider all of these challenges and solutions as researchers begin to explore the synthesis of actinide-based thin film heterostructures.

\begin{figure*}
    \centering
    \includegraphics[width=\textwidth]{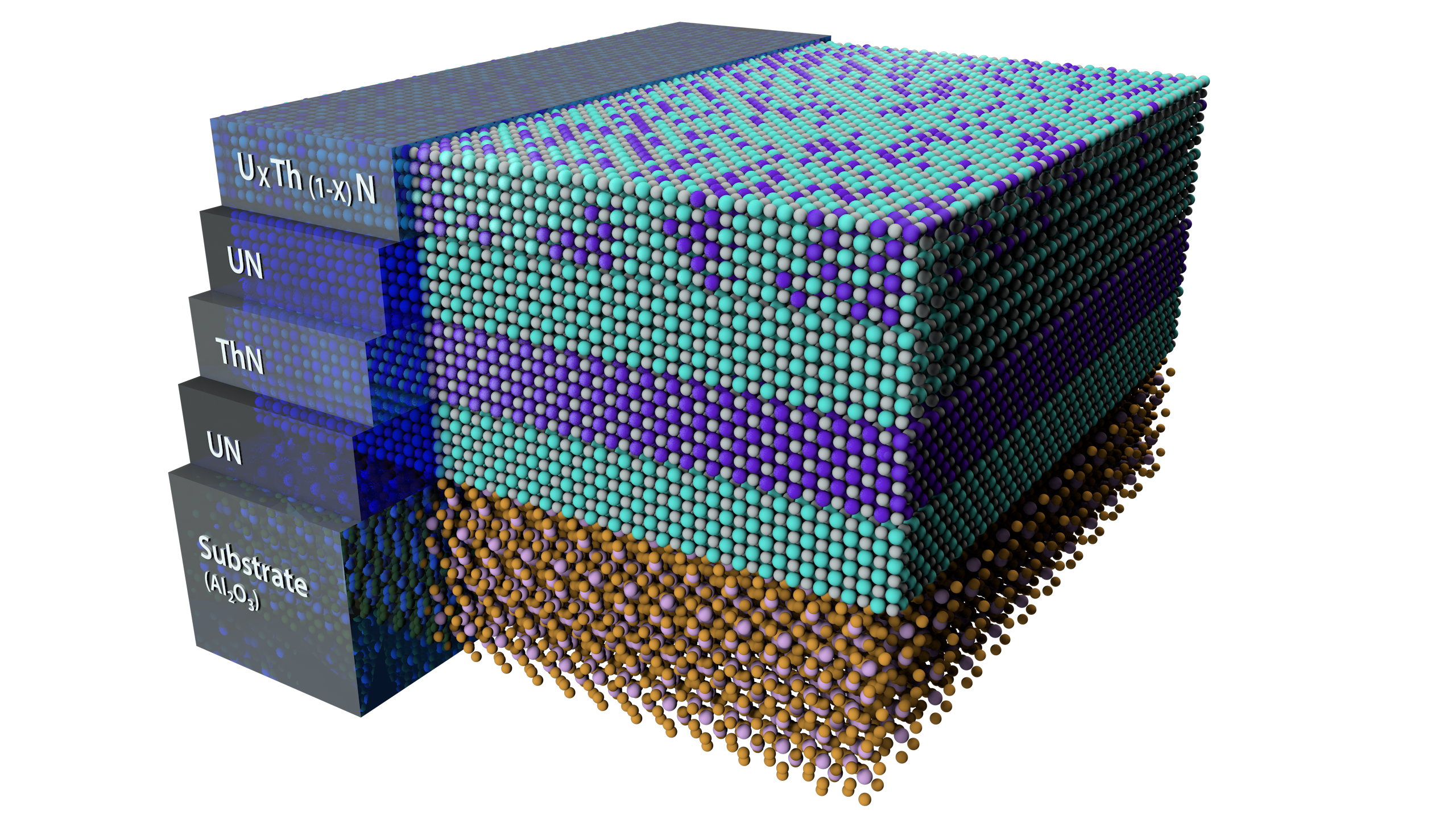}
    \caption{Sketch of a proposed heterostructure consisting of alternating UN and ThN thin films. In practice, to avoid strain relaxation, the maximum allowed thickness of each layer must be calculated from the Matthews-Blakeslee equation~\cite{matthews1974defects}.}
    \label{fig:hetero}
\end{figure*}

To fill this technological gap, the Idaho National Lab has recently commissioned an MBE system designed specifically for the growth of actinide-nitride compounds. This first-of-its-kind MBE system will permit feasibility studies of the growth of various actinide-based thin films. Proof-of-concept experiments will target the MBE growth of heterostructures based on UN and ThN~(Figure \ref{fig:hetero}). These heterostructures will enable the study of the electronic properties of these interesting 5$f$-electron heterointerfaces. Given the lattice mismatch of $\sim5.7\%$ between UN and ThN, the substrate material must be chosen carefully to allow growth of layers with reasonable thickness, while remaining below the Matthews-Blakeslee limit for strain relaxation~\cite{matthews1974defects}. 

An alternative approach is to create heterostructures based on U$_x$Th$_{1-x}$N alloys that have lower lattice mismatch to the binary end-point compounds. To this end, Table \ref{tab:mbe4} lists a ``recipe book" of actinide nitrides and candidate substrates with reasonably lattice-matched alloys. Indeed, by incorporating suitable buffer layers, some of these compound/substrate combinations have already been synthesized~\cite{bright2018epitaxial}. The hope is that by exploring the MBE growth of materials in these actinide nitride systems, their thermophysical, magnetic, and topological properties can be studied with high precision. Actinide-nitride-based systems that show promising initial results will be screened for emergent phenomena, with a view to opening a new approach to investigating the interplay between electronic correlations and topology in epitaxial thin films and heterostructures.

\scriptsize
\begin{table*}[ht!]
\caption{Crystallographic properties of UN and ThN thin films and candidate substrates.}
\label{tab:mbe4}
\begin{tabular}{|c|c|c|c|c|c|c|}
\hline
\textbf{Material} & \textbf{Group} & \textbf{$a$ (\AA)} & \textbf{$c$ (\AA)} & \textbf{Substrate} & \textbf{Substrate lattice const. (\AA)} & \textbf{Strain}  \\  [0.5ex]
\hline
UN  & Fm$\bar{3}$m & 4.88  & 4.88  &  $\langle001\rangle_{\text{UN}}\|\langle001\rangle_{\text{Al}_2\text{O}_3}$ & 4.785 & 1.96\%   \\
 & Fm$\bar{3}$m & 4.88  & 4.88  &  $\langle001\rangle_{\text{UN}}\|\langle111\rangle_{\text{Si}}$ & 3.135 & \textit{relaxed}   \\
 & Fm$\bar{3}$m & 4.88  & 4.88  &  $\langle001\rangle_{\text{UN}}\|\langle111\rangle_{\text{NaF}}$ & 4.62 & 5.84\%  \\
\hline
UN$_2$   &  Fm$\bar{3}$m & 5.31 & 5.31  & $\langle111\rangle_{\text{UN$_2$}}\|\langle001\rangle_{\text{LAO}}$ & 3.79 &\cite{rough1958constitution} \\
   &  Fm$\bar{3}$m & 5.31 & 5.31  & $\langle100\rangle_{\text{UN$_2$}}\|\langle001\rangle_{\text{ZnS}}$ & 5.41  &1.02\%\\
   &  Fm$\bar{3}$m & 5.31 & 5.31  & $\langle100\rangle_{\text{UN$_2$}}\|\langle001\rangle_{\text{GaAs}}$ & 5.3565 &0.7\%\\
   &  Fm$\bar{3}$m & 5.31 & 5.31  & $\langle100\rangle_{\text{UN$_2$}}\|\langle001\rangle_{\text{Si}}$ & 5.43 &2.26\%\\
\hline
$\alpha$-U$_2$N$_3$    &  Ia$\bar{3}$            & 10.684 & 10.684 & $\langle001\rangle_{\text{U}_2\text{N}_3}\|\langle001\rangle_{\text{Y$_2$O$_3$}}$ & 10.64 & 0.75\% \\
\hline
$\beta$-U$_2$N$_3$    &  Ia$\bar{3}$            & 3.69 & 5.826  & $\langle001\rangle_{\text{U}_2\text{N}_3}\|\langle001\rangle_{\text{CaF}_2}$ & 5.47 & {2.05\% (2:1 relation)} \\
\hline
ThN         & Fm$\bar{3}$m & 5.20  & 5.20  &  $\langle001\rangle_{\text{ThN}}\|\langle001\rangle_{\text{CaF}_2}$         &  5.47 & 5.37\% \\
         & Fm$\bar{3}$m & 5.20  & 5.20  &  $\langle001\rangle_{\text{ThN}}\|\langle001\rangle_{\text{YSZ}}$ & 5.125 & 0.76\% \\
        & Fm$\bar{3}$m & 5.20  & 5.20  &  $\langle001\rangle_{\text{ThN}}\|\langle001\rangle_{\text{Si}}$         & 5.43 & 5.00\% \\
\hline
Th$_2$N$_3$   & C$\bar{3}$m & 3.875 & 6.175  &  $\langle001\rangle_{\text{Th}_2\text{N}_3}\|\langle001\rangle_{\text{LAO}}$         & 3.79 & 2.19\% \\
\hline
Th$_3$N$_4$  & R$\bar{3}$m & 3.875  & 27.39 & $\langle001\rangle_{\text{Th}_3\text{N}_4}\|\langle001\rangle_{\text{LAO}}$ & 3.79 & 2.19\%\\
\hline
\end{tabular}
\end{table*}
\normalsize

\section{Outlook}

Recent years have seen significant advances in the field of actinide materials synthesis. Ongoing improvements in materials growth techniques are beginning to produce single-crystal samples of actinide metals and their compounds. These efforts include the synthesis of both bulk crystals and thin films, within which grain boundaries are eliminated and the density of other crystallographic defects is reduced. Syntheses using a variety of methods have been demonstrated important initial material compatibility criteria (temperature ranges, lattice constants, etc.) for reliable thin film synthesis. As a result, researchers can, for the first time, investigate thermal and electrical transport properties of actinides in samples with dramatically reduced phonon and electron scattering. In the near term, concerted efforts are being made to integrate different thin film actinides into heterostructures whose properties can be tuned through strain, confinement and other interfacial effects. The expectation is that, as a result, the subtle physical features of these materials will begin to emerge. Exotic magnetic and topological properties are expected to arise in these strongly correlated systems with 5\textit{f}-shell electrons, leading to the discovery of new physics. 

\section*{Acknowledgements}

This work was supported through the INL Laboratory Directed Research \& Development Program under U.S. Department of Energy Idaho Operations Office Contract DE-AC07-05ID14517. F.K. acknowledges the support of the Glenn T. Seaborg Institute for Actinide Science at Idaho National Laboratory. C.A.M and D.H.H. acknowledge support from the Center for Thermal Energy Transport under Irradiation (TETI), an Energy Frontier Research Center funded by the US Department of Energy, Office of Science, Office of Basic Energy Sciences. The authors would like to thank A. Tiwari at the University of Utah for useful discussions. K.D.V. acknowledges the support of his herds of guinea pigs and hamster.

\section*{References}
\bibliographystyle{unsrt}
\bibliography{aipsamp}

\end{document}